# Virtual-mask Informed Prior for Sparse-view Dual-Energy CT Reconstruction

Zini Chen [1,#], Yao Xiao [1,#], Junyan Zhang [1,*], Shaoyu Wang [2], Liu Shi [2,*], Qiegen Liu [2]

[1] School of Mathematics and Computer Sciences, Nanchang University, Nanchang 330031, China.
[2] Department of Information Engineering, Nanchang University, Nanchang 330031, China.
[#] These authors contributed to this work equally.
*Corresponding author (Email: zhangjunyan@email.ncu.edu.cn and shiliu@ncu.edu.cn).

*Abstract*—**Sparse-view sampling in dual-energy computed tomography (DECT) significantly reduces radiation dose and increases imaging speed, yet is highly prone to artifacts. Although diffusion models have demonstrated potential in effectively handling incomplete data, most existing methods in this field focus on the image domain and lack global constraints, which consequently leads to insufficient reconstruction quality. In this study, we propose a dual-domain virtual-mask informed diffusion model for sparse-view reconstruction by leveraging the high inter-channel correlation in DECT. Specifically, the study designs a virtual mask and applies it to the high-energy and low-energy data to perform perturbation operations, thus constructing high-dimensional tensors that serve as the prior information of the diffusion model. In addition, a dual-domain collaboration strategy is adopted to integrate the information of the randomly selected high-frequency components in the wavelet domain with the information in the projection domain, for the purpose of optimizing the global structures and local details. Experimental results indicated that the present method exhibits excellent performance across multiple datasets.**

# 1. Introduction

Dual-energy computed tomography (DECT) operates on a unique imaging mechanism that scans objects simultaneously with two X-rays of different energies [1-3]. In contrast to traditional single-energy counterpart, this fashion allows it to obtain abundant information under high-energy and low-energy X-ray irradiation [4]. In practical scenarios, to enhance acquisition efficiency and lower radiation dose, reducing projection views is a common strategy [5]. However, sparse sampling leads to data in-completeness, which often results in artifacts in the recon-structed images, significantly degrading the image quality [6].

Reconstructing sparse views in DECT poses a challenging inverse problem, how to improve its reconstruction quality has been a frontier area in recent years [7]. Numerous classical reconstruction methods have been put forward to enhance image quality. For instance, Long et al. [8] employed the multi-channel penalized weighted least squares (PWLS) method to estimate the data fidelity error. To boost the robustness of reconstruction, Yu et al. [9] pro-posed the prior image-constrained compressed sensing (PICCS) method. A method utilizing tensor decomposition was presented to address noise issues during reconstruction [10]. Nevertheless, these classical approaches are computationally costly, and their effectiveness is often restricted by parameters [11]. In recent years, deep learning techniques have shown remarkable feature-extraction and non-linear mapping capabilities. They are able to learn the latent image structures from data, thus significantly improving the DECT reconstruction quality [12, 13]. For example, Zhang et al. [14] introduced a novel one-step inverse generation network for sparse-view DECT, which can achieve simultaneous imaging of spectral images and materials. Additionally, a reconstruction strategy based on the generative adversarial network named DER-GAN was proposed by Xiang et al [15], which used a strip-shaped modulator placed in front of the detector to acquire dual-energy projections in a single scan, thereby enabling the reconstruction of incomplete views.

To further improve the reconstruction quality in sparse-view DECT, researchers are committed to exploring multi-domain collaborative reconstruction methods. Zhang et al. [16] combined the advantages of the image domain and the projection domain and proposed CD-Net, which enhanced the reconstruction quality through neural net-works and reconstruction operator. Similarly, Zhu et al. [17] proposed the MsDu-Nets method, which carried out DECT reconstruction using a multi-stage dual-domain network, utilized the information from the projection do-main and the image domain to reduce artifacts. Also, Wang et al. [18] constructed the DoDa-Net dual-domain bidirectional estimation network, which achieved data conversion from the projection domain to the image do-main, and optimized the image domain data using the im-age processing network. In addition, Liu et al. [19] constructed diffusion model constraint terms in the image and wavelet spaces and obtained good experimental results. Although the above-mentioned methods have achieved promising results in improving the reconstruction quality, how to make more in-depth use of the high similarity between different channels of DECT to further enhance the reconstruction quality remains a key issue worthy of further exploration [20].

To address the above-mentioned issues, inspired by the high similarity between the two energy channels in DECT, the study proposes a sparse-view reconstruction method called Virtual-mask Informed Prior (VIP-DECT). The design of the virtual mask greatly explores the correlations between the DECT channels and proposes a new infor-mation-association pattern. Furthermore, virtual-mask perturbations are applied separately in the projection domain and the wavelet domain, thus constructing high-dimensional tensors that serve as prior information for the diffusion model, enhancing its ability to capture both the global and local features of the data. During the training procedure in the wavelet domain, VIP-DECT performs wavelet transformation on the high-energy and low-energy data to obtain four components separately, and randomly selects a pair of high-frequency components from them for the above-mentioned operations. Instead of including all components in the training procedure, this strategy reduces the computational cost, enabling the sampling process to be carried out in multiple diffusion steps, which improves the convergence speed and the quality of image reconstruction, as shown in Fig. 1. The core innovations consist of the following two components:

- **High-Energy and Low-Energy Channel Information Interaction:** Virtual masks are randomly generated to perturb the data distribution, which enable the information inter-action between high-energy and low-energy channels. Based on this interaction, VIP-DECT constructs high-dimensional tensors that contain the original data, which sig-nificantly utilizes the correlations in DECT channels, providing prior information for diffusion models. The novel prior information helps the model to learn the data features more effectively.

- **Global and Local Information Collaboration in Dual-Domain:** Considering the effective utilization of global and local information, a dual-domain collaborative strategy integrating the projection domain and the wavelet do-main is adopted. The projection domain is closely related to the physical mechanism of DECT and places more emphasis on the global features. Moreover, high-frequency components of the wavelet domain are randomly selected, enabling the model to focus more on the reconstruction of local details.

The subsequent sections of the paper are organized as follows: Section II briefly introduces the relevant research. Section III provides a detailed formulation of the pro-posed method and discusses the prior learning and joint iterative reconstruction process of VIP-DECT. Section IV evaluates the performance of the proposed method in sparse-view reconstruction and compares it with other methods. Section V discusses and concludes the study.

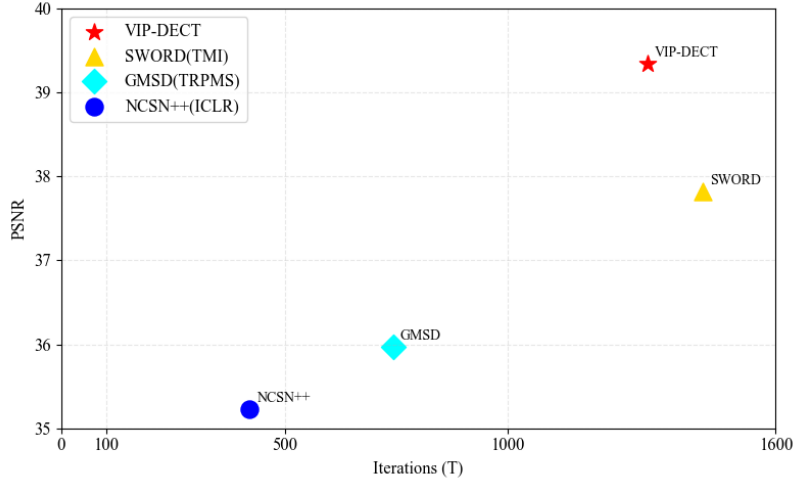

**Fig. 1.** Convergence analysis plot of NCSN++, GMSD, SWORD and VIP-DECT in terms of PSNR versus the number of iterations during sparse-view reconstruction on the head DECT dataset at the iteration numbers corresponding to the peaks.

# 2. Related Work

## 2.1 Sparse-View DECT Reconstruction

Sparse sampling in DECT is a linear measurement process. In practical applications, sparse projection data is often obtained by subsampling the projection data of DECT [21]. The forward projection process can be represented as the following discrete linear system [22]:

$$X = AI \tag{1}$$

where $X \in \mathbb{R}^{V \times D \times C}$ is the obtained projection data of DECT, $I \in \mathbb{R}^{H \times W \times C}$ represents the original DECT image, and $A$ represents the projection matrix. Specially, $V$ represents the projection views, $D$ represents the rays (detector elements) of each projection, and $H$, $W$, $C$ here represent the height, width, and channel number of $I$. Fig. 2. illustrates the process of converting image data into sparse-view projection data. Under the influence of the subsampling mask for the projection data, the standard full projection data is sparsely sampled to produce a small-sized sparse DECT, and the mathematical formulation of the linear process is as follows:

$$y = P(\Lambda)AI = P(\Lambda)X \tag{2}$$

where $y \in \mathbb{R}^{V \times D \times C}$ represents the measurement obtained after sparse sampling, $P(\Lambda)$ is a 0-1 matrix for the sparse sampling of the projection data, and its generation depends on the needed sparse views. Under sparse-view condition, the incompleteness of data causes the reconstruction problem to be ill-posed [23]. To overcome it, researchers introduce a regularization term to constrain the reconstructed images and solve for the optimal solution by minimizing a cost function that takes both data consistency and prior knowledge into account [24-26], which can be expressed as:

$$\min_{I}\left[\|y - P(\Lambda)X\|_2^2 + \lambda R(I)\right] \tag{3}$$

where the objective function consists of two data fidelity terms $\|y - P(\Lambda)x\|_2^2$, a regularization term $R(I)$ and the factor $\lambda$ used to balance these two types of terms.

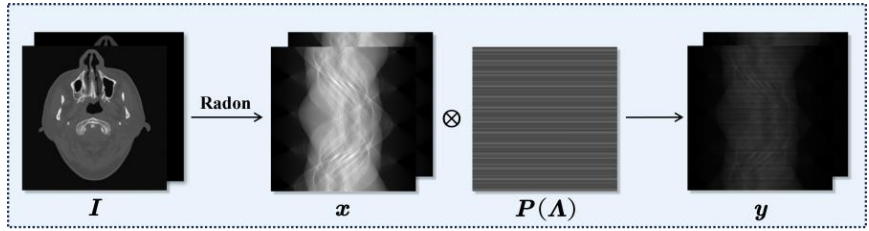

**Fig. 2.** Sparse sampling process in sparse-view CT.

## 2.2 Diffusion with Stochastic Differential Equations (SDEs)

The score-based generative diffusion model injects Gaussian noise at various levels to disrupt the data [27]. Simultaneously, it trains a deep neural network with noise as a conditioning factor. The diffusion model encompasses both forward and reverse stochastic differential equations. The forward stochastic differential equation is as follow:

$$dx = f(x,t)dt + g(t)d\omega \tag{4}$$

where $f(x,t) \in \mathbb{R}^n$ is the drift coefficient, $g(t) \in \mathbb{R}$ is the diffusion coefficient, $\omega \in \mathbb{R}^n$ represents the standard Brownian motion. In the prior learning stage, the Variance Exploding (VE) SDE is pivotal. It transforms the complex data distribution into a known prior distribution. Based on this, the forward VE-SDE can be precisely expressed as a Gaussian noise function that changes dynamically:

$$dx = \sqrt{\frac{[d(\sigma^2(t))]}{dt}}\,d\omega \tag{5}$$

where the noise function $\sigma(t)$ changes with time within the continuous-time interval $t \in [0, 1]$. In practical scenarios, calculating $\nabla_x \log p(x)$ at all time points is often extremely challenging and cannot be directly acquired. Thus, we optimize model parameters $\theta$ to train a time-dependent model. This enables us to train a time-dependent model $\nabla \hat{\log} p_\theta(x_t, t)$. The objective function is defined as:

$$L(\theta) = \mathbb{E}_{t,x_0} \big[ \| \nabla \hat{\log} p_\theta(x_t, t) - \nabla \log p(x_t) \|_2^2 \big] \tag{6}$$

where $t$ is a time variable uniformly sampled from the time interval $[0, T]$ and $\mathbb{E}$ denotes the expectation. By minimizing this objective function, the model can efficiently learn time-related score information. Once the network converges to the optimal model, it implies that the trained model can effectively approximate the true score function.

The reverse SDE is the core method for generating the original samples from noisy data [28], which can be expressed as:

$$dx = [f(x, t) - g^2(t) \nabla_x \log p_t(x)] dt + g(t) d\omega \tag{7}$$

where $\nabla_x \log p_t(x)$ is the score function. Time gradually decreases from a larger value to 0, simulating the restoration of the original data from a noisy state. As time progresses, $f(x, t)$ guides the data towards the original data distribution, while the random factors introduced by $g(t)$ help prevent the data from getting trapped in local optima. These two terms balance each other, allowing the reverse SDE to stably and effectively recover high-quality original samples from the noise.

# 3. PROPOSED METHOD

In this section, a detailed description of the proposed VIP-DECT is presented. First, the motivation underlying the VIP-DECT is articulated. Next, the core concepts of the diffusion models in the projection domain and wavelet domain are introduced. Finally, a joint iterative optimization model is presented. The detailed process of the whole training and sampling reconstruction strategy is also exhibited.

## 3.1 Motivation

DECT images consist of two channels: high ($H$)-energy and low ($L$)-energy channels, which exhibit a notable correlation. To precisely and visually confirm the correlation, the study utilizes a Region-based Structural Similarity Index (Region-based SSIM), as illustrated in Fig. 3. Inspired by the high correlation among channels in DECT, we propose a virtual mask strategy that conforms to the Gaussian distribution. As depicted in Fig. 3, randomly generated virtual masks are used to perturb the components of the original tensor, thereby obtaining perturbed tensors. Subsequently, these tensors are stacked together with the original tensor to form a high-dimensional tensor [29-31], which lays a solid foundation for efficiently and accurately performing the task of sparse-view DECT reconstruction.

Furthermore, combining the mask strategy with diffusion models offers significant advantages. Diffusion models have demonstrated great potential in generating high-quality images through the process of gradually adding noise and then reversing it [32]. The mask strategy proposed in this study uses virtual masks following the Gaussian distribution to perturb the original tensor, thus constructing a high-dimensional tensor rich in prior information. When this mask strategy is integrated with diffusion models, the prior information contained in the high-dimensional tensor can provide crucial guidance for the diffusion process, enabling it to guide the diffusion process in a more focused way and allowing the diffusion model to accurately focus on the key features in the perturbed data during processing.

To thoroughly explore the optimization potential for sparse-view DECT reconstruction, a multi-dimensional and meticulous analysis of the reconstruction process is essential. Currently, most DECT reconstruction works focus on the image domain. Despite achieving some results, image domain reconstruction often encounters is-sues like information loss when handling complex tissue structures. In contrast, applying the proposed mask strategy in the projection domain has unique advantages, as its intrinsic connection to the DECT acquisition source makes it a highly effective approach. It enables source-level data optimization and effectively constrains the re-construction process globally, leading to more accurate and reliable reconstructions. After applying the proposed mask strategy in the projection domain, during the subsequent optimization efforts, we discover that co-optimizing with the wavelet domain might further improve the reconstruction quality of sparse views [33]. Wavelet transformation can decompose an image into high ($h$)-frequency and low ($l$)-frequency components [34]. The $h$-frequency components contain abundant detail information [35], which can offset the information deficiency in sparse-view data, thus enhancing the quality and completeness of the reconstructed images. The study innovatively selects a set of $h$-frequency components randomly in the wavelet domain and implements the novel random perturbation mechanism to conduct information interaction between $H$-energy and $L$-energy channels. The incorporation of randomness empowers the model to more efficiently learn and make use of the useful information within the $h$-frequency components.

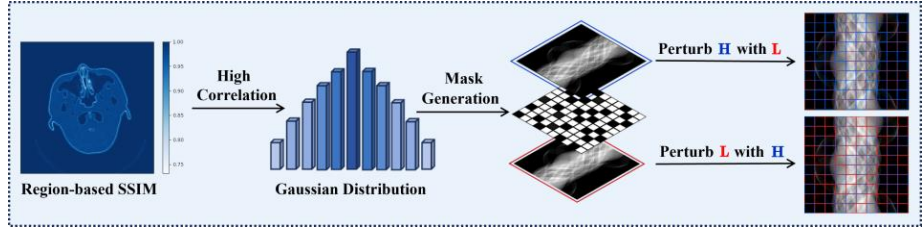

**Fig. 3.** Illustration of the channel correlation and its perturbation process. It shows the channel correlation and its perturbation process, including the correlation situation of the region-based SSIM, mask generation process, and process of using the generated mask to perturb the original tensor to obtain the perturbed tensor.

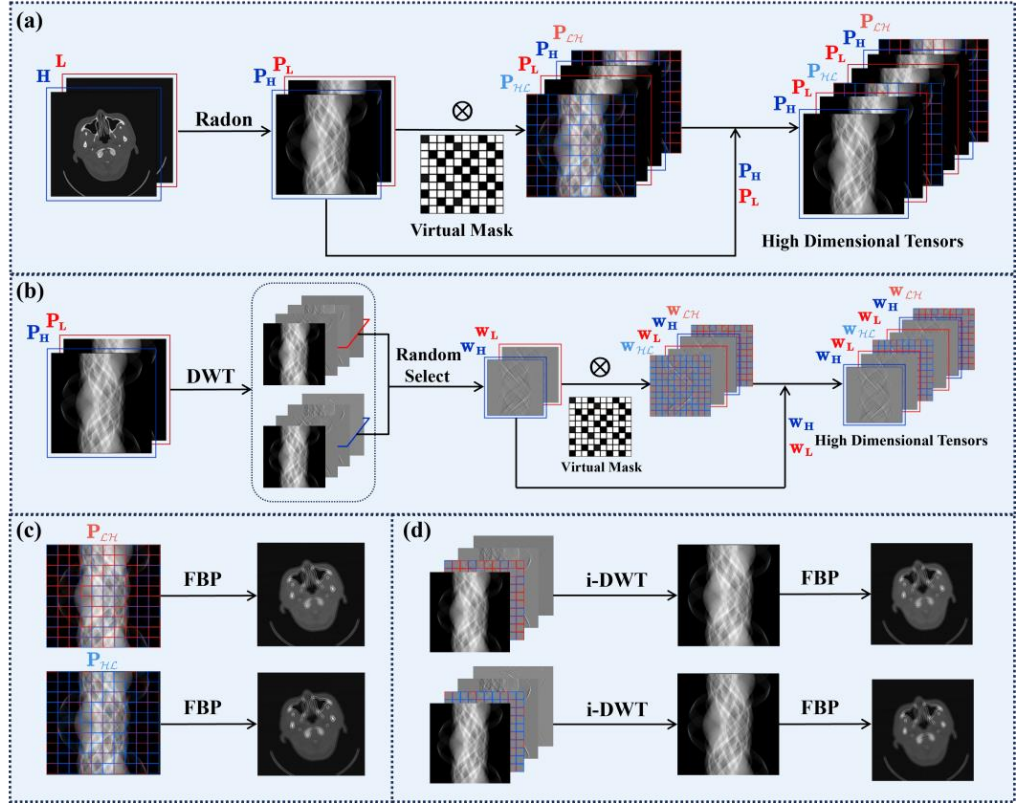

**Fig. 4.** Illustration of the virtual mask cross-energy transformation. (a) VCT process in projection domain. (b) VCT process in wavelet domain, (c) The corresponding image from the perturbated projection domain, (d) The corresponding image from the perturbated wavelet domain.

## 3.2 Domain Channel Disturbance Procedure

**1) *Virtual Mask Generation.*** In the process of generating the virtual mask, we first generate a random matrix sampled from a Gaussian distribution. Then, we apply a predefined threshold to convert the values within the matrix into a binary mask. Specifically, each element of the mask is determined according to whether the corresponding random value exceeds the threshold. If it exceeds the threshold, the element is set to 1; otherwise, it is set to 0. The generated binary mask has the same spatial dimensions as the input image data. The density of the mask is regulated by adjusting the threshold, thus achieving both randomness and controllability in the perturbation process. This approach ensures that the perturbation maintains consistency in spatial structure while introducing variability across channels, enhancing the interaction and correlation between $H$-energy and $L$-energy channels. The generated mask $M$ can be expressed as:

$$M(i,j) = \begin{cases} 1 & , \; if \; x \leq \zeta \\ 0 & , \; if \; x > \zeta \end{cases} \tag{8}$$

where $x$ is a random variable that follows a Gaussian distribution $\mathcal{N}(0,1)$, and $\zeta$ is the perturbation ratio.

2) **Virtual Mask Cross-energy Transformation.** In view of the strong correlation between $H$-energy and $L$-energy channels in DECT, the study proposes a channel perturbation method via virtual mask cross-energy transformation (VCT). As an encoding transformation strategy, VCT reencodes the data space, which constructs high-dimensional tensors by perturbing $H$-energy and $L$-energy channels using virtual masks. Concretely, we randomly generate virtual masks that are congruent with the size of the input data blocks. Subsequently, random perturbations are carried out on $H$-energy and $L$-energy channels of the original DECT. This yields two perturbed tensors: $x_{per1} = [HL, L]$ and $x_{per2} = [H, LH]$. This perturbation process is depicted in Fig. 4(a)(b), which can be expressed by the following formula:

$$VCT: \begin{cases} C_{\mathcal{HL}} = C_H \odot M + C_L \odot (1-M) \\ C_{\mathcal{LH}} = C_L \odot M + C_H \odot (1-M) \end{cases} \tag{9}$$

where $C_H$ and $C_L$ represent $H$-energy and $L$-energy tensors, $C_{\mathcal{HL}}$ and $C_{\mathcal{LH}}$ are the tensors resulting from virtual mask perturbation. $M$ represents the matrix corresponding to the virtual mask, and the symbol "$\odot$" denotes the dot-product operation. Thereafter, we stack these perturbed tensors and the original tensor in an organized manner to create a high-dimensional tensor $x_h = [H, L, HL, L, H, LH]$. As shown in Fig. 4(c)(d), it can be observed that the images obtained through back-projection after perturbation contain rich structural information. Therefore, the perturbed data, as an element of the high-dimensional tensor, can significantly enhance the ability of model to capture complex features. The formed high-dimensional tensor encompasses the characteristic changes of the original data under different perturbations, making the information to be more diverse and unique, which can provide extremely effective prior information for subsequent model training.

3) **Inverse Virtual Mask Cross-energy Transformation.** During the iterative reconstruction, to enhance the effectiveness of the encoding strategy in both the training and testing phases, the study introduces inverse VCT (i-VCT) transformation module. Unlike the VCT that constructs high-dimensional tensors by perturbing channels, the i-VCT aims to reverse this perturbation process. This reverse perturbation process is depicted in Fig. 5, which can be expressed as follows:

$$i-VCT: \begin{cases} C_{\mathcal{L}} = C_{\mathcal{H}\mathcal{L}} \odot (1-M) + C_{\mathcal{L}\mathcal{H}} \odot M \\ C_{\mathcal{H}} = C_{\mathcal{L}\mathcal{H}} \odot (1-M) + C_{\mathcal{H}\mathcal{L}} \odot M \end{cases} \tag{10}$$

where $C_{\mathcal{H}}$ and $C_{\mathcal{L}}$ represent the channel tensors obtained through the inverse virtual mask perturbation. By means of the saved virtual masks and corresponding inverse transformation methods, it can accurately restore the perturbed channels to their original channels. This process of first perturbing to construct high-dimensional interaction and then performing inverse transformation for restoration forms a closed-loop of information processing.

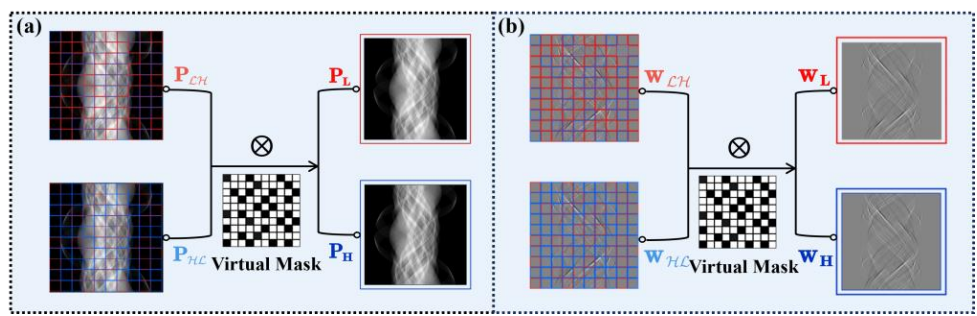

**Fig. 5.** The process of inverse mutual perturbation transformation. (a) Projection domain, (b) Wavelet domain.

## 3.3 VIP-DECT: Training Strategy

The training architecture in VIP-DECT consists of two main models. The following sections will present in-depth and detailed expositions of them, as shown in Fig. 6(a)(b) and Alg. VIP-DECT.

**1) *Projection Perturbation Model:*** The study proposes a training process that applies the VCT operations to the projection domain diffusion model via the **P**rojection **P**erturbation **M**odel (PPM). First, we map the input images to the projection domain via the Radon transform. This step converts two-dimensional image information into multi-view projection data. Then, the VCT is applied to the data, which generates high-dimensional tensors, and the process can be mathematically represented as:

$$\hat{x}_p = VCT(x_p) \tag{11}$$

where $x_p \in \mathbb{R}^{v \times d \times c}$ denotes the projection domain data, $\hat{x}_p \in \mathbb{R}^{v \times d \times c}$ represents the high-dimensional tensor resulting from the VCT. Among them, $v \times d \times c$ represents the dimensions after padding operations. Subsequently, the high-dimensional tensor, formed by combining the perturbed and original data, are used as prior information of the diffusion model. These tensors are put into a score-based network learning framework. This integration incorporates randomness and channel correlations into the training process. In the projection domain, the forward process of the VE-SDE can be expressed as:

$$dx_p = \sqrt{\frac{[d(\sigma_p^2(t))]}{dt}}\, d\omega \tag{12}$$

where $\sigma_p$ is the noise function, and the noise function $\sigma(t)$ changes with time within the continuous-time interval $t \in [0, 1]$. The objective function for this optimization is:

$$\mathcal{L}_p = \mathbb{E}_{t \sim U(0,T)}\left[\gamma_t\, \mathbb{E}_{x_{p_0} \sim p(x_{p_0})}\, \mathbb{E}_{x_{pt}|x_{p_0}}\ \| s_{\theta p}(x_{pt}, t) - \nabla_{x_{pt}} \log p_t(x_{pt}|x_{p_0}) \|_2^2\right] \tag{13}$$

where $\mathcal{L}_p$ represents the optimization objective of the projection domain, $\gamma_t$ is a positive function, and $\nabla_{x_{pt}} \log p_t(x_{pt}|x_{p_0})$ is the log-gradient of the conditional probability density function $p_t(x_{pt}|x_{p_0})$.

**2) Wavelet Perturbation Model:** To enhance the reconstruction quality of $h$-frequency details in DECT, the study proposes a model that undergoes random mask perturbation in the wavelet domain, via the **W**avelet **P**erturbation **M**odel (WPM). Specifically, first, wavelet transformation is applied to the $H$-energy and $L$-energy data of DECT images to obtain the $h$-frequency and $l$-frequency components. The mathematical expression for this decomposition process is as follows:

$$W: x_p \rightarrow w_{ll}, w_{lh}, w_{hl}, w_{hh} \tag{14}$$

where $W$ represents the single-energy wavelet transform operation, $w_{ll} \in \mathbb{R}^{\frac{v}{2} \times \frac{d}{2} \times c}$ is the $l$-frequency component, $w_{lh} \in \mathbb{R}^{\frac{v}{2} \times \frac{d}{2} \times c}$, $w_{hl} \in \mathbb{R}^{\frac{v}{2} \times \frac{d}{2} \times c}$, and $w_{hh} \in \mathbb{R}^{\frac{v}{2} \times \frac{d}{2} \times c}$ are the $h$-frequency components. Next, we randomly select one $h$-frequency component corresponding to both $H$-energy and $L$-energy data from the acquired $h$-frequency components, which can be expressed as:

$$x_{w_H} = Random(w_{H-lh}, w_{H-hl}, w_{H-hh}) \tag{15}$$

$$x_{w_L} = Random(w_{L-lh}, w_{L-hl}, w_{L-hh}) \tag{16}$$

where $x_{w_H} \in \mathbb{R}^{\frac{v}{2} \times \frac{d}{2} \times \frac{c}{2}}$ and $x_{w_L} \in \mathbb{R}^{\frac{v}{2} \times \frac{d}{2} \times \frac{c}{2}}$ represent $H$-energy and $L$-energy tensors selected from the $h$-frequency component, $x_w \in \mathbb{R}^{\frac{v}{2} \times \frac{d}{2} \times c}$ is a combination of the $x_{w_H}$ and $x_{w_L}$ channels. Then, the VCT is applied to $x_w$ as in the projection domain, and the process can be expressed as:

$$\hat{x}_w = VCT(x_w) \tag{17}$$

where $\hat{x}_w \in \mathbb{R}^{\frac{y}{2} \times \frac{d}{2} \times 3c}$ represents the high-dimensional tensor resulting from the VCT. Finally, we use $x_w$ as prior information, which is fed into a score-based network to learn the data distribution. In the wavelet domain, the forward process of the VE-SDE is expressed as:

$$dx_w = \sqrt{\frac{[d(\sigma_w^2(t))]}{dt}} \, d\omega \tag{18}$$

where $\sigma_w$ is the noise function. The objective function for this optimization is:

$$\mathcal{L}_w = \mathbb{E}_{t \sim U(0,T)} \left[ \gamma(t) \mathbb{E}_{x_{w0} \sim p(x_{w0})} \mathbb{E}_{x_{wt}|x_{w0}} \left\| s_{\theta w}(x_{wt}, t) - \nabla_{x_{wt}} \log p_t(x_{wt}|x_{w0}) \right\|_2^2 \right] \tag{19}$$

where $\mathcal{L}_w$ represents the optimization objective of the wavelet domain.

## 3.4 VIP-DECT: Iterative Reconstruction

1) *Cascade Reconstruction:* A joint-optimization model is developed, which further enhances the reconstruction quality of DECT images by jointly optimizing the PPM and WPM. During the iterative process, the two models update alternately and complement one another. This effectively combines the perturbation information from the projection domain and the wavelet domain, thereby improving the accuracy of image structure and the restoration of $h$-frequency details. To enable the model to accurately recover the original samples from noisy data, the perturbation models related to the reverse SDE in the projection domain and wavelet domain need to be jointly optimized through alternating updates, which can be formalized into the following two steps:

$$dx_p = [f_p(x_p, t) - g_p^2(t) \nabla_{x_p} \log p_t(x_p)] dt + g_p(t) d\omega_p \tag{20}$$

$$dx_w = [f_w(x_w, t) - g_w^2(t) \nabla_{x_w} \log p_t(x_w)] dt + g_w(t) d\omega_w \tag{21}$$

The alternating optimization strategy between the projection domain and the wavelet domain can be formalized as:

$$x_p^{t-1} = \mathcal{I}_p(x_p^{t-\frac{1}{2}}, M) \; ; \; x_w^{t-1} = \mathcal{I}_w(x_w^{t-\frac{1}{2}}, M) \tag{22}$$

where $x_p^{t-1}$ and $x_w^{t-1}$ represent the results of the projection domain and the wavelet domain after the step $t$ perturbation, prediction-correction update. $M$ is the virtual mask used for VCT and i-VCT. $\mathcal{I}_p$ and $\mathcal{I}_w$ represent the comprehensive transformation operations in the projection domain and the wavelet domain.

At this stage, the projection domain PC sampler, the wavelet domain PC sampler, data-consistency constraints, the VCT and i-VCT steps are executed alternately. This optimization strategy between the projection domain and the wavelet domain ensures that the structures in both domains and $h$-frequency information are fully restored and enhanced. The detail process is shown in Fig. 6(c) and Alg. 1.

**2) *Projection Domain Update:*** First, the data in projection domain is perturbed via the VCT, which is based on the Eq. (10), to form high-dimensional tensors. Then, we use a predictor to update the projection domain to obtain the intermediate result $\hat{x}_p^{t-\frac{1}{2}} \in \mathbb{R}^{v \times d \times 3c}$. The formula for using the predictor in projection domain can be expressed as follows:

$$\hat{x}_p^{t-1} = \hat{x}_p^{t-\frac{1}{2}} + (\delta_t^2 - \delta_{t-1}^2)(s_{\theta p}(x_p^{t-\frac{1}{2}}, t) + \sqrt{\delta_t^2 - \delta_{t-1}^2} z_p \quad (23)$$

where $\delta_{t-1}$ is a function that monotonically increases with time, and $z_p \sim \mathcal{N}(0,1)$ is a Gaussian-distributed random variable in projection domain. Subsequently, data consistency constraints are introduced. The data consistency constraints aim to ensure the consistency between the current projection-domain data and the original data, which can be expressed mathematically as:

$$\hat{x}_p^{t-\frac{1}{2}} = (1 - P(\Lambda'))\hat{x}_p^{t-\frac{1}{2}} + P(\Lambda')x \quad (24)$$

Next, optimize $\hat{x}_p^{t-1}$ using the corrector, which can be expressed as follows:

$$\hat{x}_p^{t-1} = \hat{x}_p^{t-1} + \epsilon_{t-1} s_{\theta p}(x_p^{t-1}, t) + \sqrt{2\epsilon_{t-1}} z_p' \quad (25)$$

Then, apply data consistency constraints to $\hat{x}_p^{t-1}$ obtained after optimization again. Subsequently, conduct the i-VCT, as per Eq. (11), to restore the image structure.

**3) *Projection Domain Update:*** Similar operations are performed in wavelet domain. First, apply the DWT to the obtained $\hat{x}_p^{t-1}$ to generate $h$-frequency and $l$-frequency components. For the $w_{lh}$, $w_{hl}$, and $w_{hh}$ corresponding to $H$-energy and $L$-energy data, perform the VCT one by one on the three groups of $h$-frequency components corresponding to the $H$-energy and $L$-energy data according to Eq. (10) to form the corresponding high-dimensional tensors. Then, input $\hat{x}_w^{t-1} \in \mathbb{R}^{\frac{v}{2} \times \frac{d}{2} \times 3c}$ of the above three groups of tensors into the predictor one by one, and the process can be formalized as follows:

$$\hat{x}_w^{t-1} = \hat{x}_w^{t-\frac{1}{2}} + \left(\delta_t^2 - \delta_{t-1}^2\right)\left(s_{\theta_w}\left(x_w^{t-\frac{1}{2}}, t\right) + \sqrt{\delta_t^2 - \delta_{t-1}^2}\, z_w \right) \tag{26}$$

Subsequently, just as in projection domain update, data consistency constraints are introduced for the above three groups of tensors that have passed through the predictor, and the process can be formalized as follows:

$$x_w^{t-\frac{1}{2}} = W\left[\left(1 - P(\Lambda')\right)\hat{x}_p^{t-\frac{1}{2}} + P(\Lambda')x\right] \tag{27}$$

Next, optimize $\hat{x}_w^{t-1}$ of the three groups of tensors with a corrector one by one, which can be formalized as follows:

$$\hat{x}_w^{t-1} = \hat{x}_w^{t-1} + \epsilon_{t-1}s_{\theta_w}\left(x_w^{t-1}, t\right) + \sqrt{2\epsilon_{t-1}}\, z_w' \tag{28}$$

Then, after the perturbation and optimization of corrector in the wavelet domain, data consistency constraints and the i-VCT based on Eq. (11) are applied again. Next, combine the transformed $\hat{x}_w^{t-1}$ with $w_{ll}$ corresponding to $H$-energy and $L$-energy data to obtain a complete set of data in wavelet domain. Then, perform the i-DWT to convert the data back to the projection domain and enter the next iteration.

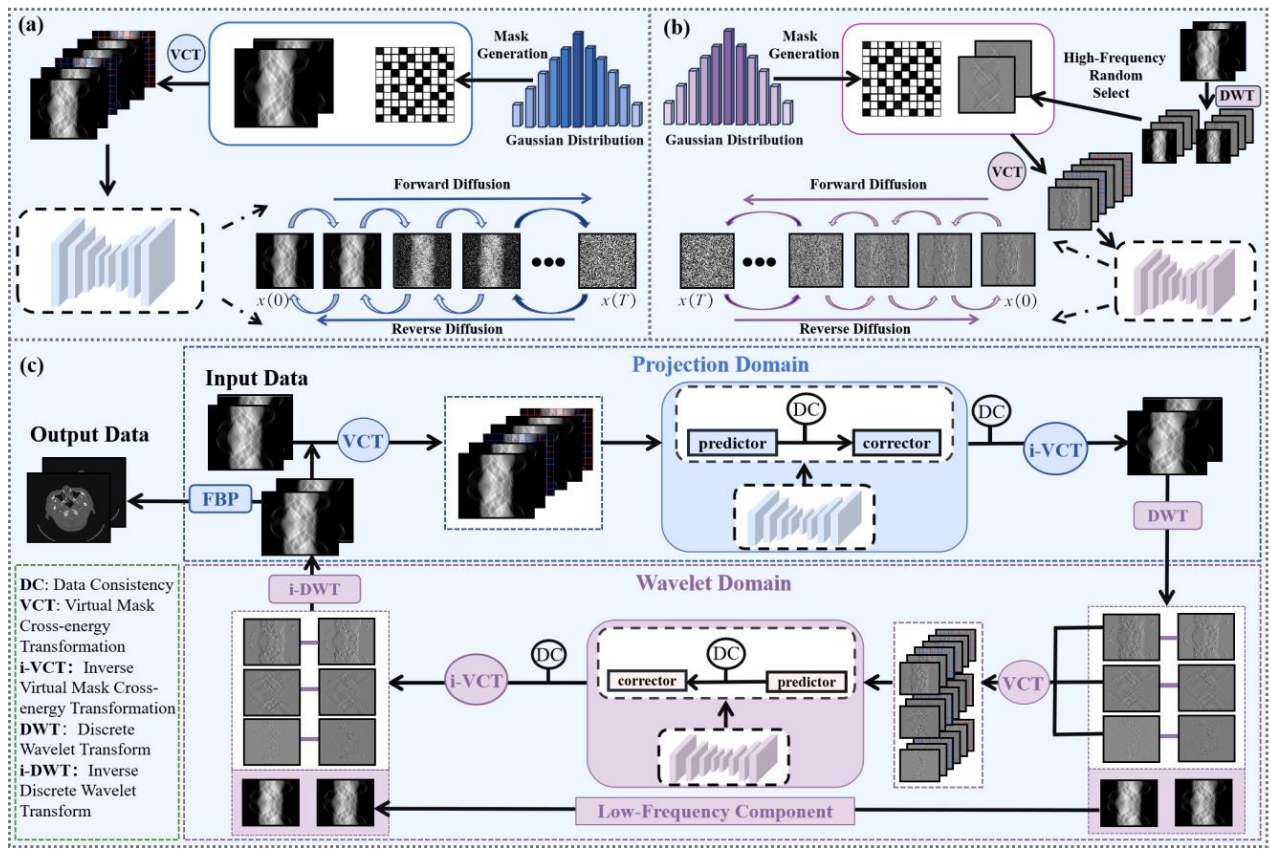

**Fig. 6.** The training stage and iterative reconstruction stage of the proposed VIP-DECT method. (a) Training scheme of PPM, (b) Training scheme of WPM, (c) Reconstruction stage.

| **Algorithm 1** |
| :--- |
| **PPM training stage** |
| **Dataset:** Several DECT image data $x$. |
| 1: Projection transform $x_p$; |
| 2: $x'_p = VCT(x_p)$; |
| 3: Train on $x'_p$. |
| **WPM training stage** |
| **Dataset:** Several DECT image data $x$. |
| 1: Get $w_{lh}$, $w_{hl}$, and $w_{hh}$; |
| 2: Randomly select $x_w$; |
| 3: $x'_w = VCT(x_w)$; |
| 4: Train on $x'_w$. |
| **VIP-DECT reconstruction stage** |
| **Initialize:** $N$. |
| 1: Radon transform $x_p$; |
| 2: For $i = N-1$ to $0$ do: |
| 3:    $\hat{x}_p = VCT(x_p)$; |
| 4:    Update $\hat{x}_p$ via Eq. (23) (Predictor); |
| 5:    Update $\hat{x}_p$ via Eq. (25) (Corrector); |
| 6:    Update $\hat{x}_p$ via Eq. (24) (DC); |
| 7:    $x_p = i - VCT(\hat{x}_p)$; |
| 8:    Wavelet transformation and select $h$-frequency as $x_w$; |
| 9:    $\hat{x}_w = VCT(x_w)$; |
| 10:   Update $\hat{x}_w$ via Eq. (26) (Predictor); |
| 11:   Update $\hat{x}_w$ via Eq. (28) (Corrector); |
| 12:   Update $\hat{x}_w$ via Eq. (27) (DC); |
| 13:   $x_w = i - VCT(\hat{x}_w)$; |
| 14:   Add low-frequency to $\hat{x}'_w$; |
| 15:   $x_p = i - DWT(x_w)$; |
| 16: End for. |
| 17: $x_{rec} = FBP(x_p)$ for $n = 1, 2, ....$; |
| 18: Return $x_{rec}$. |

# 4. Experiments

In this section, the detailed experimental settings are introduced and the experimental results obtained from multiple datasets are presented. Related code is available at: https://github.com/yqx7150/VIP-DECT.

## 4.1 Experiment Setup

**1) *Datasets:*** The head dataset used in the study consists of 1505 head CT image slices sized 512×512. The dataset was acquired on February 20, 2023, from the Radiology Department of the 988th Hospital of the Chinese People's

Liberation Army, with the aid of the Siemens SOMATOM Spectral CT imaging system, which were collected from the Radiology Department of the 988 People's Hospital. Among these images, 1400 were used for training, and the remaining 105 were used for sampling. Simulated energy spectra for overlapping spectral ranges of 0-80 kVp and 0-140 kVp were generated using the SpekCalc software [36]. The energy spectra are applied to the materials through the following equation:

$$x_j = -\ln \sum_{E=0}^{E_m} S_j(E) \exp\left[ -\sum_{k=1}^{K} \varepsilon_k \tau_k(E) \right] \tag{29}$$

where $x_j = 1, 2$ represent $H$-energy and $L$-energy images, $S_j(E)$ and $\tau_k(E)$ are the selected spectra and attenuation coefficients for different materials at the corresponding energies. This allows us to obtain the noise-free $H$-energy and $L$-energy images. During the scanning process of the simulation dataset, the distance from the X-ray source to the object (Source-to-Object Distance, SOD) was 1000 mm, the distance from the X-ray source to the detector (Source-to-Detector Distance, SDD) was 1500 mm, and the detector size was 0.204 mm.

To further validate the generalization of the proposed algorithm, the study selects the mouse thoracic cavity as real data for validation. The anesthetized mouse was scanned using an advanced MARS multi-energy CT system, which was conducted in accordance with animal welfare ethics review guidelines and was approved by the Ethics Committee of the Li Ka Shing Faculty of Medicine, The University of Hong Kong, for animal experiments (20 February 2023). The dataset includes two different energy ranges: 7-70 kVp and 70-120 kVp. During scanning, the SOD was 156 mm, SDD was 256 mm, and pixel size was 0.110 mm.

The study extracted 30, 60 and 90 views data from the full projection for sparse-view CT reconstruction. To simulate and generate projection data for the fan-beam CT reconstruction task, we used the Siddon ray-driven algorithm [37]. The distance from the light source and the detector to the rotation center is 40 cm, the detector width is 41.3 cm, with 720 detector elements in total, and all projection views are evenly distributed over a 360-degree range.

2) **Parameter Setting:** Experiments of this study are conducted on a high-performance workstation (Tesla V100-PCIE-16 GB), using the Operator Discretization Library (ODL)[38] and PyTorch within a Python environment. In the experiment, the Adam optimizer is used, with a learning rate set to 0.0002, to train the PPM and WPM, and the Kaiming initialization method is applied to initialize the network weights [39]. Based on our experience, the experiment set $\sigma_{\min}$ to be 0.01 and $\sigma_{\max}$ to be 378. As time progresses, the noise intensity gradually increases. In the recon-struction phase, the number of iterations is set to 2000. For each execution of the prediction and correction process, annealed Langevin dynamics are applied. Furthermore, the signal-to-noise ratio is given by $SNR = 0.075$. The perturbation rate $\zeta$ of the virtual masks in VCT and i-VCT during the training and sampling processes is empirically set to 2.81, and the relevant Gaussian distribution has a mean $\mu = 1$ and a standard deviation $\sigma = 0$.

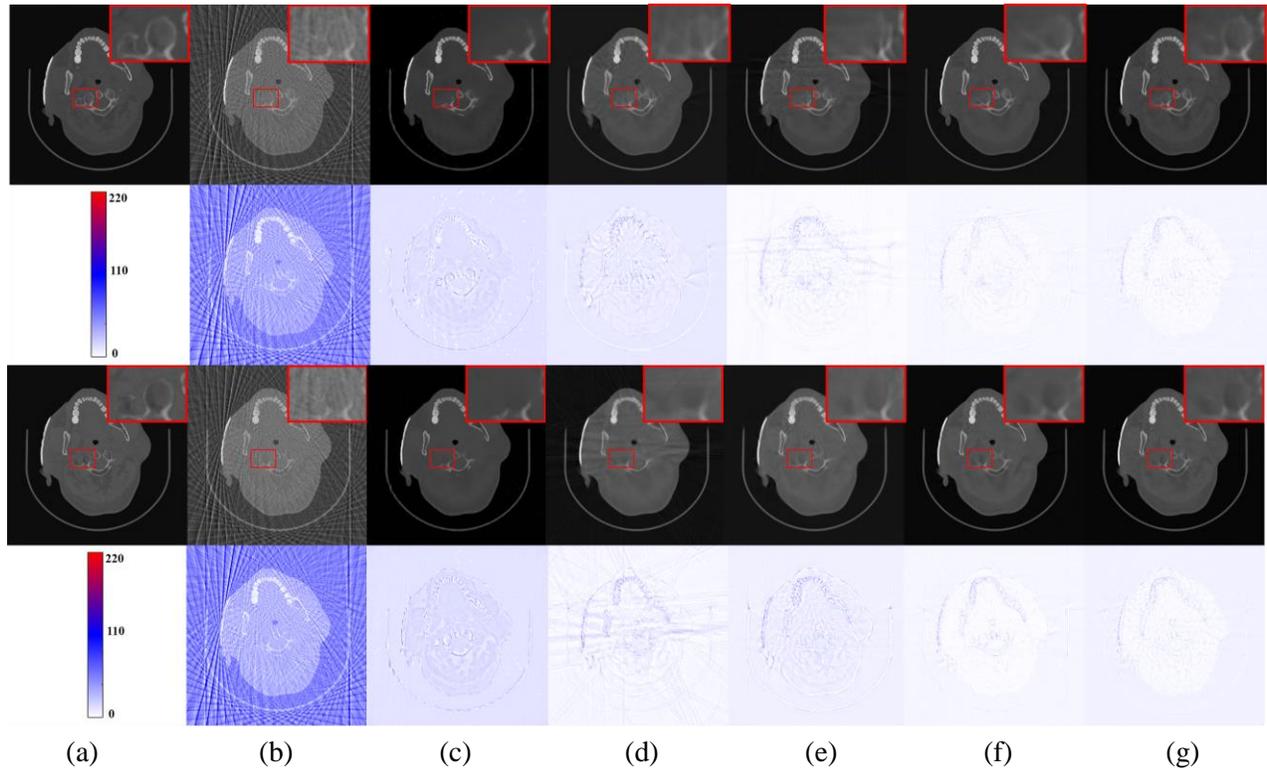

(a)  (b)  (c)  (d)  (e)  (f)  (g)

**Fig. 7.** Reconstruction results from 60 views on the head DECT dataset. The first and third rows show $H$-energy and $L$-energy reconstructed images, while the second and fourth rows show residuals between reference and reconstructed images, with the display window set to the range of [0, 220] HU. (a) The reference image versus the images reconstructed by (b) FBP, (c) FBPConvNet, (d) NCSN++, (e) GMSD, (f) SWORD, and (g) VIP-DECT.

3) *Evaluation Metrics:* To quantitatively evaluate the performance of the model, the study calculates performance metrics for the reconstructed $H$-energy and $L$-energy images, by selecting three widely-used standard metrics: PSNR, SSIM, MSE. Notably, in this study, the calculation of these indicators is not directly compared with the original real images, the real images are first subjected to the Radon transform, and then the data resulting from the application of FBP are utilized. Through a comprehensive evaluation of these three indicators, we can comprehensively understand the performance during reconstruction and quantify the differences and similarities between the reconstructed images and the true images reconstructed by projection and FBP.

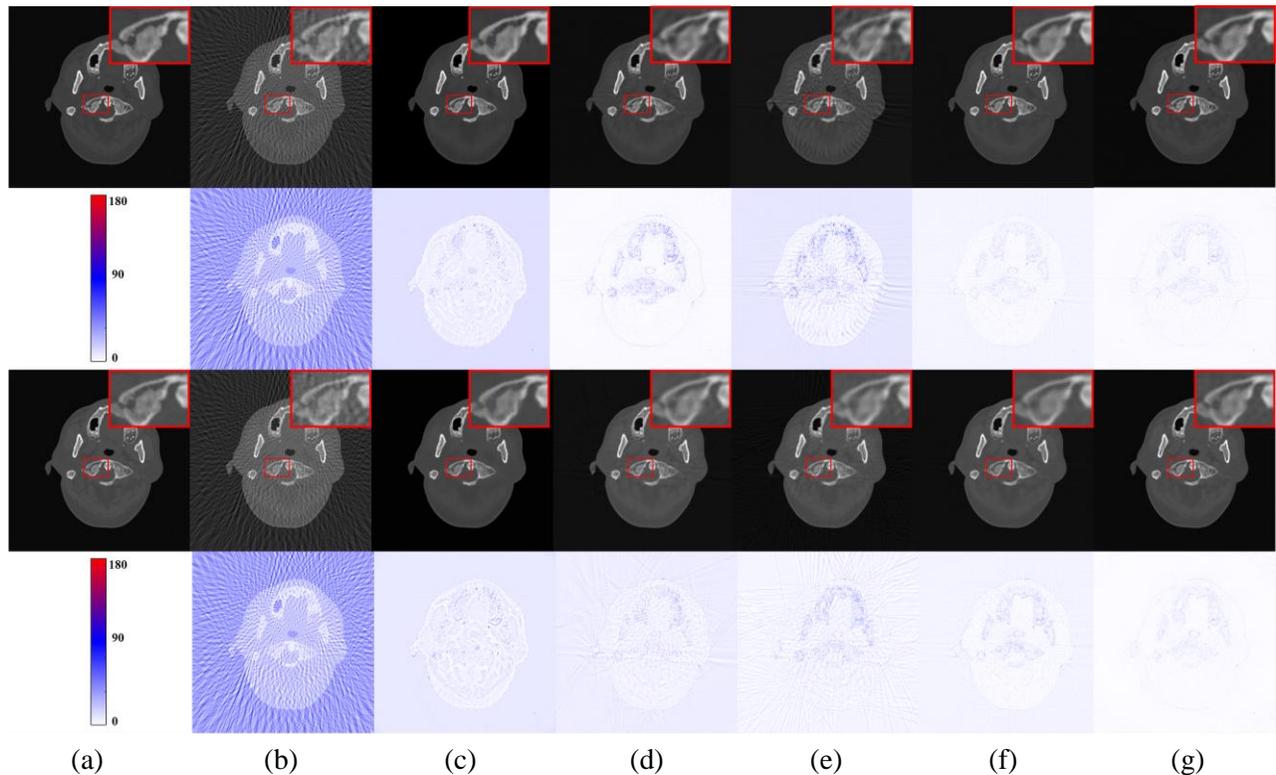

|         |         |         |         |         |         |         |
| (a)     | (b)     | (c)     | (d)     | (e)     | (f)     | (g)     |

**Fig. 8.** Reconstruction results from 90 views on the head DECT dataset. The first and third rows show $H$-energy and $L$-energy reconstructed images, while the second and fourth rows show residuals between reference and reconstructed images, with the display window set to the range of [0, 180] HU. (a) The reference image versus the images reconstructed by (b) FBP, (c) FBPConvNet, (d) NCSN++, (e) GMSD, (f) SWORD, and (g) VIP-DECT.

## 4.2 Experimental Comparison

In this section, the experimental results of evaluating different methods on the aforementioned datasets are presented.

**1) *Head DECT Dataset:*** The study chooses multiple image slices from the head dataset, where these slices exhibit a variety of structural and textural features. Then, experiments are carried out using 30, 60 and 90 views, and meanwhile, a comparative analysis of several methods is conducted. Among these methods, FBP is a classic method [40], FBPConvNet combines FBP with deep-learning techniques [41], NCSN++ is based on the score-based diffusion model [27], GMSD is a sinogram domain inspired diffusion model [28], and SWORD incorporates $h$-frequency and $l$-frequency information in the wavelet domain[35]. It is worth noting that the main innovation of GMSD lies in the utilization of projection domain, while NCSN++ is originally used in the image domain. However, in this comparative experiment, NCSN++ is trained and tested using data from the projection domain.

TABLE I

PSNR/SSIM/MSE ($10^{-3}$) VALUES OF $H$-ENERGY IMAGES RECONSTRUCTED BY DIFFERENT METHODS AT 30, 60 AND 90 VIEWS FROM THE HEAD DECT DATASET.

| Algorithm | | FBP [40] | FBPConvNet [41] | NCSN++ [27] | GMSD [28] | SWORD [35] | **VIP-DECT** |
|---|---|---|---|---|---|---|---|
| View | 30 | 20.81/0.1711/8.350 | 28.79/0.9211/1.338 | 29.95/0.9326/1.137 | 28.05/0.8980/1.848 | 31.63/0.9291/0.688 | **33.30/0.9618/0.513** |
| | 60 | 23.73/0.2511/4.252 | 36.50/0.9742/0.225 | 35.71/0.9798/0.280 | 35.27/0.9657/0.333 | 39.03/0.9848/0.134 | **39.77/0.9886/0.109** |
| | 90 | 27.93/0.3938/1.624 | 38.28/0.9916/0.156 | 42.65/0.9933/0.056 | 40.54/0.9816/0.091 | 44.67/0.9949/0.035 | **44.87/0.9956/0.036** |

TABLE II

PSNR/SSIM/MSE ($10^{-3}$) VALUES OF $L$-ENERGY IMAGES RECONSTRUCTED BY DIFFERENT METHODS AT 30, 60 AND 90 VIEWS FROM THE HEAD DECT DATASET.

| Algorithm | | FBP [40] | FBPConvNet [41] | NCSN++ [27] | GMSD [28] | SWORD [35] | **VIP-DECT** |
|---|---|---|---|---|---|---|---|
| View | 30 | 20.73/0.1770/8.481 | 30.47/0.9520/0.923 | 28.48/0.8389/1.448 | 30.736/0.9229/0.889 | 28.91/0.9094/1.312 | **33.43/0.9573/0.486** |
| | 60 | 23.74/0.2586/4.270 | 35.10/0.9599/0.351 | 35.23/0.9477/0.302 | 35.97/0.9751/0.260 | 37.82/0.9844/0.167 | **39.34/0.9875/0.12** |
| | 90 | 27.95/0.4013/1.615 | 38.15/0.9884/0.165 | 41.33/0.9897/0.075 | 40.46/0.9906/0.092 | 43.17/0.9945/0.052 | **44.58/0.9955/0.037** |

To visually illustrate the reconstruction results of each method, Fig. 7 and 8 present the reconstruction results for Slice 1 with 60 views and Slice 2 with 90 views, along with their corresponding residual maps. It is obviously found that reconstruction results using the FBP often exhibit obvious artifacts. As for the FBPConvNet, although it has some improvements compared to FBP, it still falls short in accurately restoring the details of the original images. From the residual maps corresponding to NCSN++ and GMSD, it is evident that artifacts remain prominent, indicating that they are significantly affected by the loss of sparse-view data. The SWORD results in the loss of some fine details that should be present in the original images during the reconstruction process, leading to an incomplete representation of the images. In contrast, the proposed VIP-DECT method stands out remarkably. It can not only effectively eliminate artifacts and recover more details, making the reconstruction results closer to the ground truth, but also demonstrate excellent performance under different numbers of views. When 60 views are used, the images generated by VIP-DECT already present rich details and complete structures. As the number of views increases to 90, the reconstruction quality improves significantly, revealing extremely fine details and being close to the ground truth, fully demonstrating its generalization ability and outstanding performance as an unsupervised method.

To further validate the visual observations from a quantitative perspective, the study presents the reconstruction results in Table I and II, along with a comparative analysis of the average PSNR, SSIM, and MSE values of the selected slices. The optimal values of the reconstructed images corresponding to different views are emphasized in bold. Ex-periment results show that, similar to the conclusions drawn from qualitative analysis, VIP-DECT shows superior performance compared to the compared reconstruction methods across all views presented in the experiment. Particularly noteworthy is that the gap between $H$-energy and $L$-energy indicators in terms of various reconstruction metrics of VIP-DECT is relatively small, which thoroughly validates the efficient exploitation of the high correlation among DECT channels, as well as the effectiveness and superiority in the proposed method.

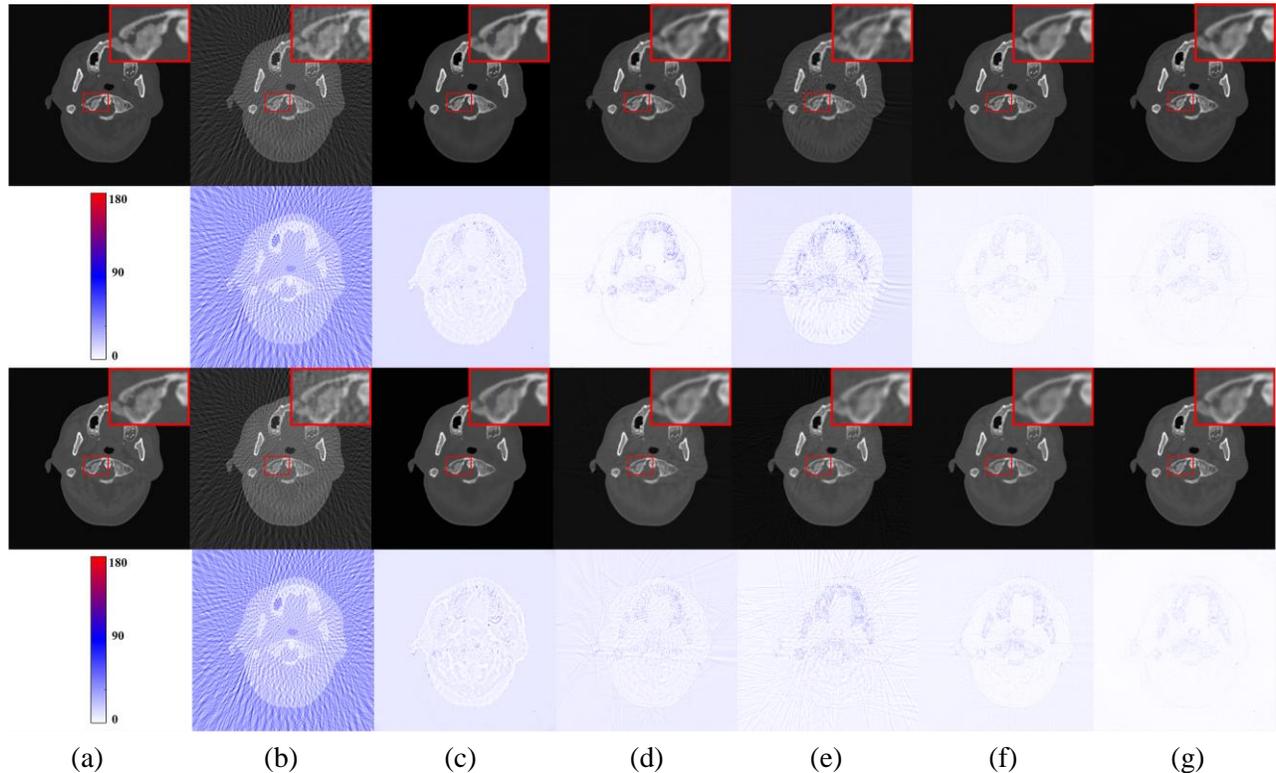

(a)        (b)        (c)        (d)        (e)        (f)        (g)

**Fig. 9.** Reconstruction results from 60 views on the mouse thoracic DECT dataset. The first and third rows show $H$-energy and $L$-energy reconstructed images, while the second and fourth rows show residuals between reference and reconstructed images, with the display window set to the range of $[0, 180]$ HU. (a) The reference image versus the images reconstructed by (b) FBP, (c) FBPConvNet, (d) NCSN++, (e) GMSD, (f) SWORD, and (g) VIP-DECT.

*2) **Mouse Thoracic DECT Dataset:*** To validate the generalization and robustness of the proposed method, the study extracts prior knowledge from the head DECT dataset and evaluates the performance of model on the mouse thoracic DECT dataset. The key evaluation metrics under different sparse-view conditions are presented in Table III and IV respectively. The experimental results indicate that VIP-DECT performs remarkably well across multiple evaluation metrics. Notably, in scenarios with fewer views, it achieves higher-quality reconstructions compared to other methods.

Fig. 9. presents $H$-energy and $L$-energy image reconstruction results with 60 views, along with their corresponding residual maps. In FBP, severe artifacts emerge in the reconstructed images. Although methods such as FBPConvNet, NCSN++, and GMSD have enhanced the image quality to a certain extent, due to the deficiency of sparse angular data,

the artifacts remain challenging to eliminate. While the SWORD can relatively effectively remove the artifacts, when compared with the method proposed in this study, there are certain missing details in the reconstructed images. Thus, the experimental results demonstrate that the VIP-DECT has a clear advantage in sparse-view reconstruction, particularly in achieving a favorable balance in reconstructing both $H$-energy and $L$-energy images.

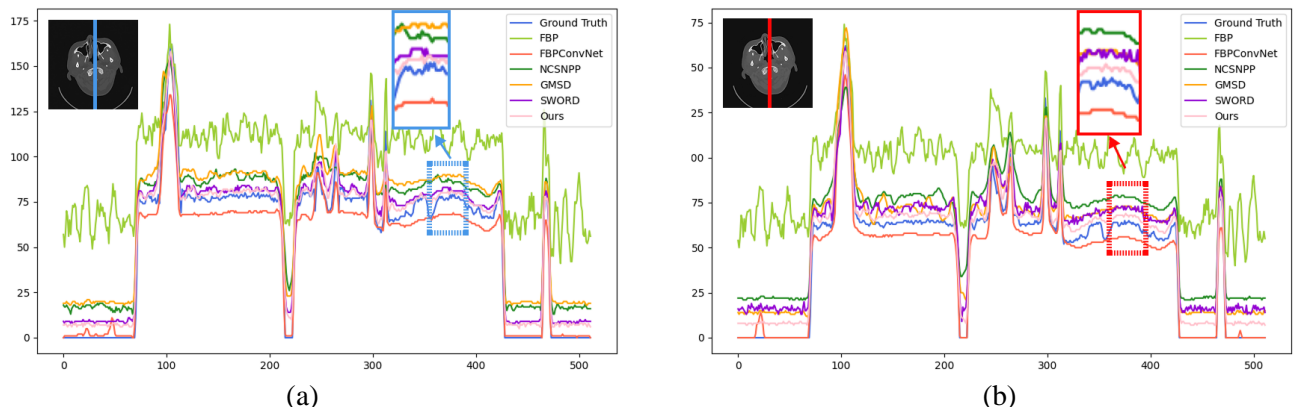

**Fig. 10.** Comparison of profile analyses for reconstruction results of different methods in sparse-view reconstruction on the head DECT dataset. Sub-figures (a) and (b) show the numerical curves of reconstruction for different profiles in $H$-energy and $L$-energy channels. Curves of different colors denote different reconstruction methods.

TABLE III

PSNR/SSIM/MSE ($10^{-3}$) VALUES OF $H$-ENERGY IMAGES RECONSTRUCTED BY DIFFERENT METHODS AT 30, 60 AND 90 VIEWS IN MOUSE THORACIC DECT DATASET.

| Algorithm | | FBP [40] | FBPConvNet [41] | NCSN++ [27] | GMSD [28] | SWORD [35] | **VIP-DECT** |
|---|---|---|---|---|---|---|---|
| | 30 | 23.14/0.2312/4.853 | 28.19/0.7399/1.518 | 29.68/0.9048/1.076 | 30.43/0.8946/0.905 | 31.69/0.9177/0.678 | **34.10/0.9404/0.389** |
| View | 60 | 27.32/0.3986/1.852 | 31.49/0.8437/0.709 | 36.14/0.9639/0.243 | 35.70/0.9387/0.269 | 36.54/0.9543/0.222 | **38.32/0.9712/0.147** |
| | 90 | 31.31/0.6014/0.740 | 32.60/0.8913/0.549 | 39.22/0.9746/0.120 | 38.68/0.9648/0.135 | 39.25/0.9749/0.119 | **40.35/0.9786/0.092** |

TABLE IV

PSNR/SSIM/MSE ($10^{-3}$) VALUES OF $L$-ENERGY IMAGES RECONSTRUCTED BY DIFFERENT METHODS AT 30, 60 AND 90 VIEWS IN MOUSE THORACIC DECT DATASET.

| Algorithm | | FBP [40] | FBPConvNet [41] | NCSN++ [27] | GMSD [28] | SWORD [35] | **VIP-DECT** |
|---|---|---|---|---|---|---|---|
| | 30 | 22.99/0.2406/5.019 | 26.71/0.7846/2.132 | 28.42/0.8476/1.439 | 28.79/0.8594/1.320 | 32.41/0.9126/0.574 | **33.89/0.9430/0.409** |
| View | 60 | 27.43/0.4110/1.805 | 31.72/0.8778/0.673 | 37.49/0.9694/0.178 | 36.29/0.9578/0.235 | 38.19/0.9687/0.152 | **38.96/0.9772/0.127** |
| | 90 | 31.53/0.6096/0.702 | 34.55/0.9157/0.351 | 40.96/0.9836/0.080 | 40.91/0.9825/0.081 | **41.15**/0.9821/**0.077** | 41.07/**0.9845**/0.078 |

## 4.3 Profile Lines Analysis

To further assess the performance of the proposed method in the sparse-view reconstruction of DECT, Fig. 10. illustrates the numerical curves of the reconstruction results from different methods along specific profiles, which are compared with the ground-truth simultaneously. The two profiles in the figure are indicated by blue and red lines in the schematic representations of $H$-energy and $L$-energy images in the upper-left corner.

As evident from the profile analysis results, in some regions, the curve of the traditional method FBP deviates substantially from the ground-truth curve, indicating certain limitations of this method in sparse-view reconstruction. Comparative methods can fit the ground-truth well at some positions but still exhibit discrepancies in other regions. In contrast, the curve of VIP-DECT is closer to the ground-truth curve at multiple key positions, which implies that our method can more accurately restore the true information of the images in sparse-view DECT reconstruction.

## 4.4 Convergence Analysis

To further validate the convergence of VIP-DECT, Fig. 11. displays the intermediate samples related to PSNR and SSIM values of the reconstruction results under 60 views. The samples at different reconstruction steps embedded in the figures are associated with the curves via yellow arrows, intuitively presenting the progress of image reconstruction at specific iteration steps. From these two sets of images, it can be seen that VIP-DECT exhibits a stable convergence trend in both $H$-energy and $L$-energy DECT sparse-view reconstruction. It can effectively improve the quality of reconstructed images at different iteration stages, further confirming its effectiveness and convergence characteristics in DECT reconstruction tasks.

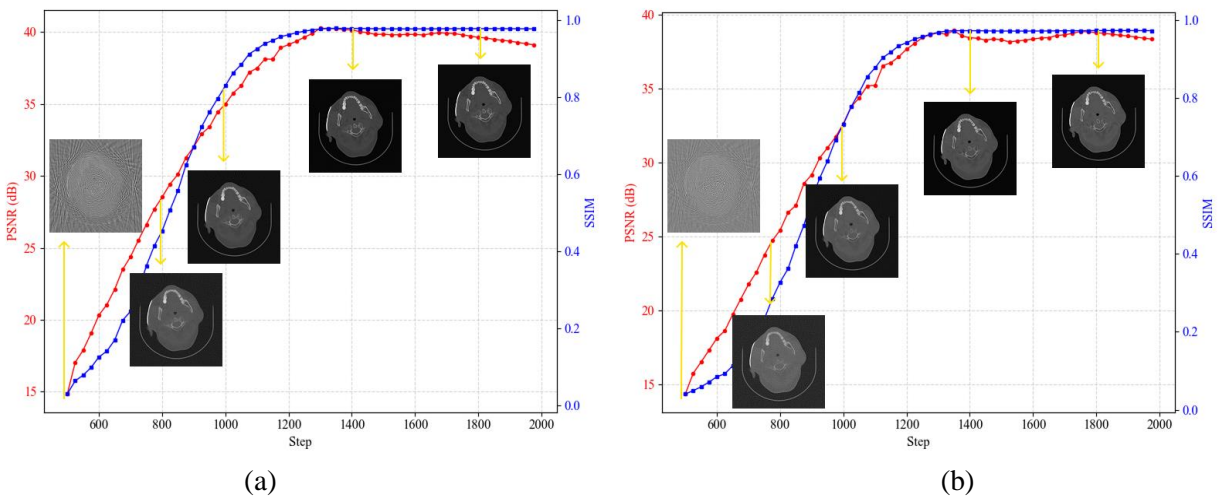

(a)                                        (b)

**Fig. 11.** PSNR and SSIM values and intermediate samples in the iterative process of (a) $H$-energy and (b) $L$-energy DECT.

## 4.5 Ablation Study

To evaluate the effectiveness of each component, we explored the effects of mask-related parameters on the recon-struction results and conducted ablation experiments on each method, ensuring the comprehensiveness and accuracy of the experiments.

**1)** ***Different Perturbation Ratios in VIP-DECT:*** According to Eq. (9), in the sparse-view DECT reconstruction, the mask perturbation ratio is one of the key factors affecting the reconstruction quality. The study sets a series of perturbation ratios of different magnitudes (2.00, 2.81, and 3.00) to conduct experiments. The goal is to explore their impacts on the performance of VIP-DECT and find an optimal perturbation ratio for model optimization. The $H$-energy reconstruction results are recorded in Table V, and the $L$-energy reconstruction results are presented in Table VI.

TABLE V

PSNR/SSIM/MSE ($10^{-3}$) RESULTS OF $H$-ENERGY RECONSTRUCTION ON THE HEAD DECT DATASET UNDER DIFFERENT PERTURBATION RATIOS.

| View | | 30 | 60 | 90 |
|---|---|---|---|---|
| | 2.00 | 33.16/0.9607/0.519 | 39.57/0.9878/0.113 | 44.25/0.9950/0.043 |
| Ratio | 2.81 | 33.30/**0.9618**/0.513 | **39.77/0.9886/0.109** | **44.87/0.9956/0.036** |
| | 3.00 | **33.59**/0.9583/**0.472** | 39.43/0.9885/0.122 | 44.68/0.9953/0.035 |

TABLE VI

PSNR/SSIM/MSE ($10^{-3}$) RESULTS OF $L$-ENERGY RECONSTRUCTION ON THE HEAD DECT DATASET UNDER DIFFERENT PERTURBATION RATIOS.

| View | | 30 | 60 | 90 |
|---|---|---|---|---|
| | 2.00 | 33.26/0.9570/0.495 | 39.17/0.9860/0.125 | 44.39/0.9952/0.038 |
| Ratio | 2.81 | 33.43/**0.9573**/0.486 | **39.34/0.9875/0.120** | **44.58/0.9955/0.037** |
| | 3.00 | **33.63**/0.9552/**0.462** | 39.02/0.9870/0.132 | 44.52/**0.9956**/0.039 |

Research has shown that alterations in the perturbation ratio endow the model with distinct characteristics. As the perturbation ratio rises from a low value, the model output manifests local data irregularities, which impacts image details. For example, at a low perturbation ratio of 2.0, the reconstructed image has a low PSNR, high MSE, and a sub-optimal SSIM value, indicating poor structural similarity. Conversely, when the perturbation ratio declines from a high value, it triggers irregular global data changes, affecting the overall quality of the reconstructed image. For instance, when the perturbation ratio is 3.0, the comprehensive performance metrics do not reach their optimal levels. Through extensive experimentation and data analysis, it has been determined that, within the scope of this experiment, a perturbation ratio of 2.81 yields the best overall performance across all metrics for the reconstructed image. This implies that this perturbation ratio enables the model to perform optimally in sparse-view DECT reconstruction, facilitating high-quality image reconstruction. Notably, given the potential disparities in data characteristics and distribution patterns among different datasets, diverse perturbation amplitudes may be needed to achieve optimal reconstruction outcomes. Future research will focus on precisely tailoring the perturbation amplitude to different datasets, aiming to enhance the generalization ability and performance of the model.

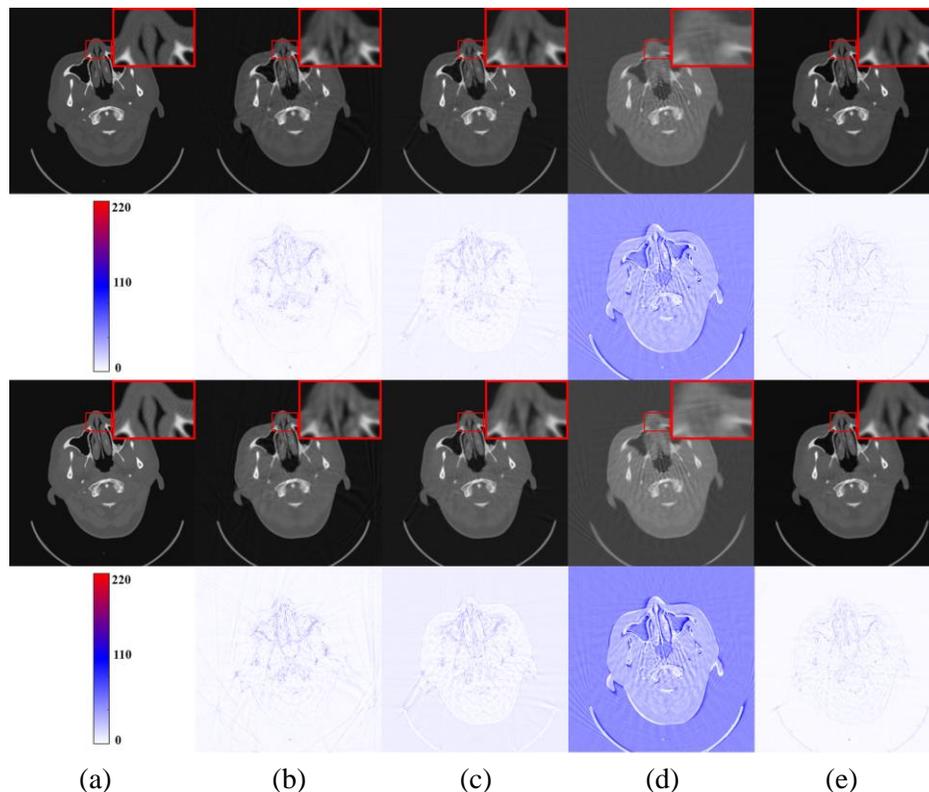

(a)         (b)         (c)         (d)         (e)

**Fig. 12.** Reconstruction results from 60 views on the head DECT dataset: The first and third rows show $H$-energy and $L$-energy reconstructed images, while the second and fourth rows show residuals between reference and reconstructed images, with the display window set to [0, 220] HU. (a) Reference image, (b) NCSN++, (c) PPM, (d) WPM, (e) VIP-DECT.

    2) *Specific Functions of Key Components in VIP-DECT:* In the experiments, VIP-DECT is decomposed into the PPM and WPM. Moreover, the NCSN++ is introduced as a baseline for comparison. The corresponding quantitative results are presented in Table VII and VIII respectively. It can be seen that VIP-DECT is significantly superior to its individual component, which further confirmed its superiority in the application of sparse-view DECT reconstruction. To further illustrate the impact of model combination, Fig. 12. presents the qualitative results of reconstruction at 60 views. The qualitative analysis shows that PPM has a relatively prominent effect in removing artifacts, while the WPM has a strong ability to restore image details. VIP-DECT combines the artifact-suppression ability of PPM and the detail-restoration ability of WPM, making full use of the complementary advantages of the two, and significantly im-

proving the quality of image reconstruction.

TABLE VII
ABLATION STUDY RESULTS OF PSNR/SSIM/MSE $(10^{-3})$ FOR $H$-ENERGY RECONSTRUCTION ON THE HEAD DECT DATASET.

| | Model | NCSN++ | PPM | WPM | **VIP-DECT** |
|---|---|---|---|---|---|
| | 30 | 31.17/0.9157/0.806 | 31.66/0.9465/0.767 | 24.29/0.7715/3.866 | **33.30/0.9618/0.513** |
| View | 60 | 36.17/0.9666/0.243 | 37.60/0.9819/0.176 | 27.36/0.8504/1.926 | **39.77/0.9886/0.109** |
| | 90 | 40.88/0.9866/0.083 | 42.02/0.9918/0.067 | 30.31/0.9003/0.978 | **44.87/0.9956/0.036** |

TABLE VIII
ABLATION STUDY RESULTS OF PSNR/SSIM/MSE $(10^{-3})$ FOR $L$-ENERGY RECONSTRUCTION ON THE HEAD DECT DATASET.

| | Model | NCSN++ | PPM | WPM | **VIP-DECT** |
|---|---|---|---|---|---|
| | 30 | 31.17/0.9068/0.801 | 31.80/0.9421/0.726 | 24.29/0.7678/3.867 | **33.43/0.9573/0.486** |
| View | 60 | 36.09/0.9605/0.250 | 37.20/0.9802/0.193 | 27.35/0.8473/1.922 | **39.34/0.9875/0.120** |
| | 90 | 40.61/0.9842/0.089 | 41.68/0.9912/0.071 | 30.27/0.8989/0.981 | **44.58/0.9955/0.037** |

# 5. Discussion and Conclusion

This study presented the VIP-DECT, which was based on a unique collaborative optimization framework and deeply integrates virtual mask technology into the diffusion model. By skillfully combining the PPM and WPM, the virtual mask can fully realize its effectiveness in different transformation domains, thereby thoroughly exploring the inherent correlation between the $H$-energy and $L$-energy channels of DECT and significantly enhancing inter-channel information interaction. During the experimental validation phase, using the simulated head DECT dataset, VIP-DECT significantly outperformed other comparison methods in several key metrics, demonstrating its outstanding stability. When tested on a real mouse thoracic DECT dataset, the model showed excellent generalization ability without the need for additional training. Experimental results clearly indicated that VIP-DECT has a significant edge over traditional baseline methods. From a practical application perspective, the VIP-DECT is expected to reduce patient radiation dose while substantially improving diagnostic accuracy, showcasing its strong capabilities in sparse-view reconstruction. However, there is still room for improvement in the time efficiency of the proposed method. Due to the inherent characteristics of the diffusion model and the dual-domain collaborative work involved in this method, the computation time is relatively long. In the future, in-depth research can be conducted on the structure of the diffusion model, and attempts can be made to simplify the unnecessary complex layers or operations in the model, thereby reducing the overall computation time.

# Acknowledgements

This work was supported in part by National Natural Science Foundation of China under 62122033 and Key Research and Development Program of Jiangxi Province under 20212BBE53001.

# References

[1]   W. Van Elmpt, G. Landry, M. Das, and F. Verhaegen, "Dual energy CT in radiotherapy: current applications and future outlook," *Radiotherapy and Oncology,* vol. 119, no. 1, pp. 137-144, 2016.

[2]   S. Aran, L. Daftari Besheli, M. Karcaaltincaba, R. Gupta, E. J. Flores, and H. H. Abujudeh, "Applications of dual-energy CT in emergency radiology," *American journal of roentgenology,* vol. 202, no. 4, pp. W314-W324, 2014.

[3]   T. R. Johnson, "Dual-energy CT: general principles," *American Journal of Roentgenology,* vol. 199, no. 5_supplement, pp. S3-S8, 2012.

[4]   R. Ghasemi Shayan, M. Oladghaffari, F. Sajjadian, and M. Fazel Ghaziyani, "Image quality and dose comparison of single‐energy CT (SECT) and dual‐energy CT (DECT)," *Radiology research and practice,* vol. 2020, no. 1, p. 1403957, 2020.

[5]   J. Di, J. Lin, L. Zhong, K. Qian, and Y. Qin, "Review of sparse-view or limited-angle CT reconstruction based on deep learning," *Laser & Optoelectronics Progress,* vol. 60, no. 8, p. 0811002, 2023.

[6]   E. J. Candès, J. Romberg, and T. Tao, "Robust uncertainty principles: Exact signal reconstruction from highly incomplete frequency information," *IEEE Transactions on information theory,* vol. 52, no. 2, pp. 489-509, 2006.

[7]   H. Kudo, T. Suzuki, and E. A. Rashed, "Image reconstruction for sparse-view CT and interior CT—introduction to compressed sensing and differentiated backprojection," *Quantitative imaging in medicine and surgery,* vol. 3, no. 3, p. 147, 2013.

[8]   Y. Long and J. A. Fessler, "Multi-material decomposition using statistical image reconstruction for spectral CT," *IEEE transactions on medical imaging,* vol. 33, no. 8, pp. 1614-1626, 2014.

[9]   Z. Yu, S. Leng, Z. Li, and C. H. McCollough, "Spectral prior image constrained compressed sensing (spectral

PICCS) for photon-counting computed tomography," *Physics in Medicine & Biology,* vol. 61, no. 18, p. 6707, 2016.

[10] D. Hu *et al.*, "SISTER: Spectral-image similarity-based tensor with enhanced-sparsity reconstruction for sparse-view multi-energy CT," *IEEE Transactions on Computational Imaging,* vol. 6, pp. 477-490, 2019.

[11] Y. S. Han, J. Yoo, and J. C. Ye, "Deep residual learning for compressed sensing CT reconstruction via persistent homology analysis," *arXiv preprint arXiv:1611.06391,* 2016.

[12] T. Lyu, W. Zhu, Y. Zhang, W. Zhao, J. Yang, and G. Wang, "Deep learning methods in dual energy CT imaging," in *Deep Learning for Advanced X-ray Detection and Imaging Applications*: Springer, 2024, pp. 43-72.

[13] W. Mustafa *et al.*, "Sparse-view spectral CT reconstruction using deep learning," *arXiv preprint arXiv:2011.14842,* 2020.

[14] X. Zhang, L. Li, S. Wang, N. Liang, A. Cai, and B. Yan, "One-step inverse generation network for sparse-view dual-energy CT reconstruction and material imaging," *Physics in Medicine & Biology,* vol. 69, no. 14, p. 145012, 2024.

[15] J. Xiang *et al.*, "DER-GAN: Dual-energy recovery GAN for conebeam CT," *IEEE Transactions on Computational Imaging,* vol. 10, pp. 28-42, 2023.

[16] Y. Zhang *et al.*, "CD-Net: Comprehensive domain network with spectral complementary for DECT sparse-view reconstruction," *IEEE Transactions on Computational Imaging,* vol. 7, pp. 436-447, 2021.

[17] G. Zhu *et al.*, "Multi-Stage dual-domain networks with informative prior and self-augmentation for dual-energy limited-angle CT reconstruction," *IEEE Transactions on Instrumentation and Measurement,* 2024.

[18] Y. Wang *et al.*, "One half-scan dual-energy CT imaging using the dual-domain dual-way estimated network (DoDa-Net) model," *Quantitative Imaging in Medicine and Surgery,* vol. 12, no. 1, p. 653, 2022.

[19] J. Liu *et al.*, "DECT sparse reconstruction based on hybrid spectrum data generative diffusion model," *Computer Methods and Programs in Biomedicine,* vol. 261, p. 108597, 2025.

[20] C. H. McCollough, S. Leng, L. Yu, and J. G. Fletcher, "Dual-and multi-energy CT: principles, technical approaches, and clinical applications," *Radiology,* vol. 276, no. 3, pp. 637-653, 2015.

[21] F. Peyrin and K. Engelke, "CT imaging: Basics and new trends," in *Handbook of Particle Detection and Imaging*: Springer, 2021, pp. 1173-1215.

[22] H. Kim and K. Champley, "Differentiable forward projector for X-ray computed tomography," *arXiv preprint*

*arXiv:2307.05801,* 2023.

[23] T. Okamoto, T. Ohnishi, and H. Haneishi, "Artifact reduction for sparse-view CT using deep learning with band patch," *IEEE Transactions on Radiation and Plasma Medical Sciences,* vol. 6, no. 8, pp. 859-873, 2022.

[24] S. Kabri *et al.*, "Convergent data-driven regularizations for ct reconstruction," *Communications on Applied Mathematics and Computation,* pp. 1-27, 2024.

[25] W. Xia, Z. Yang, Z. Lu, Z. Wang, and Y. Zhang, "RegFormer: A local–nonlocal regularization-based model for sparse-view CT reconstruction," *IEEE Transactions on Radiation and Plasma Medical Sciences,* vol. 8, no. 2, pp. 184-194, 2023.

[26] M. K. Ng, H. Shen, E. Y. Lam, and L. Zhang, "A total variation regularization based super-resolution reconstruction algorithm for digital video," *EURASIP Journal on Advances in Signal Processing,* vol. 2007, pp. 1-16, 2007.

[27] Y. Song, J. Sohl-Dickstein, D. P. Kingma, A. Kumar, S. Ermon, and B. Poole, "Score-based generative modeling through stochastic differential equations," *arXiv preprint arXiv:2011.13456,* 2020.

[28] B. Guan *et al.*, "Generative modeling in sinogram domain for sparse-view CT reconstruction," *IEEE Transactions on Radiation and Plasma Medical Sciences,* 2023.

[29] A. Cichocki *et al.*, "Tensor decompositions for signal processing applications: From two-way to multiway component analysis," *IEEE signal processing magazine,* vol. 32, no. 2, pp. 145-163, 2015.

[30] K. Makantasis, A. D. Doulamis, N. D. Doulamis, and A. Nikitakis, "Tensor-based classification models for hyperspectral data analysis," *IEEE Transactions on Geoscience and Remote Sensing,* vol. 56, no. 12, pp. 6884-6898, 2018.

[31] Y. Zhang, X. Mou, G. Wang, and H. Yu, "Tensor-based dictionary learning for spectral CT reconstruction," *IEEE transactions on medical imaging,* vol. 36, no. 1, pp. 142-154, 2016.

[32] X. Ai *et al.*, "RED: Residual estimation diffusion for low-dose PET sinogram reconstruction," *Medical Image Analysis,* p. 103558, 2025.

[33] Z. Zhou, T. Liu, B. Yu, Y. Gong, L. Shi, and Q. Liu, "Physics-informed deepCT: Sinogram wavelet decomposition meets masked diffusion," *arXiv preprint arXiv:2501.09935,* 2025.

[34] S. G. Mallat, "A theory for multiresolution signal decomposition: the wavelet representation," *IEEE*

*transactions on pattern analysis and machine intelligence,* vol. 11, no. 7, pp. 674-693, 1989.

[35] K. Xu, S. Lu, B. Huang, W. Wu, and Q. Liu, "Stage-by-stage wavelet optimization refinement diffusion model for sparse-view CT reconstruction," *IEEE Transactions on Medical Imaging,* vol. 43, no. 10, pp. 3412-3424, 2024.

[36] G. Poludniowski, G. Landry, F. Deblois, P. M. Evans, and F. Verhaegen, "SpekCalc: a program to calculate photon spectra from tungsten anode x-ray tubes," *Physics in Medicine & Biology,* vol. 54, no. 19, p. N433, 2009.

[37] R. L. Siddon, "Fast calculation of the exact radiological path for a three‐dimensional CT array," *Medical physics,* vol. 12, no. 2, pp. 252-255, 1985.

[38] J. Adler, H. Kohr, and O. Öktem, "Operator discretization library (ODL)," *Zenodo,* 2017.

[39] K. He, X. Zhang, S. Ren, and J. Sun, "Delving deep into rectifiers: Surpassing human-level performance on imagenet classification," in *Proceedings of the IEEE international conference on computer vision*, 2015, pp. 1026-1034.

[40] D. J. Brenner and E. J. Hall, "Computed tomography—an increasing source of radiation exposure," *New England journal of medicine,* vol. 357, no. 22, pp. 2277-2284, 2007.

[41] K. H. Jin, M. T. McCann, E. Froustey, and M. Unser, "Deep convolutional neural network for inverse problems in imaging," *IEEE transactions on image processing,* vol. 26, no. 9, pp. 4509-4522, 2017.